# Revolutionizing Medical Data Sharing Using Advanced Privacy Enhancing Technologies: Technical, Legal and Ethical Synthesis

James Scheibner[1] Jean Louis Raisaro[2,3], Juan Ramón Troncoso-Pastoriza[4], Marcello Ienca[1], Jacques Fellay[2,5,6], Effy Vayena[1] and Jean-Pierre Hubaux[4]

[1] Health Ethics and Policy Laboratory, Department of Health Sciences and Technology, ETH Zürich, Zürich, Switzerland
[2] Precision Medicine Unit, Lausanne University Hospital, Lausanne, Switzerland
[3] Data Science Group, Lausanne University Hospital, Lausanne, Switzerland
[4] Laboratory for Data Security, School of Computer and Communication Sciences, EPFL, Lausanne, Switzerland
[5] School of Life Sciences, EPFL, Lausanne, Switzerland
[6] Host-Pathogen Genomics Laboratory, Swiss Institute of Bioinformatics, Lausanne, Switzerland

## Abstract

*Multisite medical data sharing is critical in modern clinical practice and medical research. The challenge is to conduct data sharing that preserves individual privacy and data usability. The shortcomings of traditional privacy-enhancing technologies mean that institutions rely on bespoke data sharing contracts. These contracts increase the inefficiency of data sharing and may disincentivize important clinical treatment and medical research. This paper provides a synthesis between two novel advanced privacy-enhancing technologies (PETs): Homomorphic Encryption and Secure Multiparty Computation (defined together as Multiparty Homomorphic Encryption or MHE). These PETs provide a mathematical guarantee of privacy, with MHE providing a performance advantage over separately using HE or SMC. We argue MHE fulfills legal requirements for medical data sharing under the General Data Protection Regulation (GDPR) which has set a global benchmark for data protection. Specifically, the data processed and shared using MHE can be considered anonymized data. We explain how MHE can reduce the reliance on customized contractual measures between institutions. The proposed approach can accelerate the pace of medical research whilst offering additional incentives for healthcare and research institutes to employ common data interoperability standards.*

## Keywords

# Introduction

The current biomedical research paradigm has been characterized by a shift from intra-institutional research towards multiple collaborating institutions operating at an inter-institutional, national or international level for multisite research projects. However, despite the apparent breakdown of research barriers, there remain differences between ethical and legal requirements at all jurisdictional levels [1]. There are numerous organizational strategies that have been used to resolve these issues, particularly for international academic consortia.

For example, the International Cancer Genome Consortium (ICGC) endeavors to amass cancer genomes paired with non-cancerous sequences in a cloud environment, known as the Pan Cancer Analysis of Whole Genomes (PCAWG). The ICGC's data access compliance office (DACO) were unable to establish an international cloud under the PCAWG because of conflicts between United States and European Union data privacy laws [2]. These conflicts will be likely exacerbated with the Court of Justice of the European Union (CJEU) invalidating the US-EU Privacy Shield agreement. This decision will prevent private research organizations from transferring personal data from the EU to the US without organizational safeguards (Case C-311/2018, *Schrems II*). In addition, the COVID-19 pandemic has made sharing data for clinical trials and research imperative. However, a series of COVID-19 papers retracted due to data unavailability emphasizes the need for data sharing to encourage oversight [3]. Further, within the EU there is the potential for differences in how countries regulate the processing of health-related personal data [4]. Finally, given study restrictions it may be impossible to share data between institutions or jurisdictions [5]. Although reforms to EU data protection law have been proposed to encourage scientific data sharing,[6] at present the best available solutions remain contractual and technological measures.

In this paper, we describe how traditional data-sharing approaches relying on conventional privacy-enhancing technologies (PETs) are limited by various regulations governing medical use and data sharing. We describe two novel PETs, homomorphic encryption and secure multiparty computation, that extend the capacity of researchers to conduct privacy-preserving multisite research. We then turn to analyze the effects of regulation on using these novel PETs for medical and research data sharing. In particular, we argue these PETs guarantee anonymity as defined under the European Union (EU) General Data Protection Regulation (GDPR) and are therefore key enablers for medical data sharing. We focus on the GDPR as it currently represents a global benchmark in data protection regulations. We argue using these technologies can reduce the reliance on customized data sharing contracts. The use of standardized agreements for multiparty processing of data in concert with PETs technologies can reduce the bottleneck on research. Finally, we turn to address how these novel PETs can be integrated within existing regulatory frameworks to encourage increased data sharing whilst preserving data privacy.

# Privacy and security issues of current medical data-sharing models

Before examining novel PETs, it is necessary to examine the main models for exchanging medical data for research purposes and the limitations of conventional privacy protection mechanisms that are currently used to reduce the risk of re-identification. We synthesize the data-sharing models into three categories and analyze their main technological issues (Figure 1).

**Centralized model: trusted dealer**
The centralized model requires medical sites (i.e., data providers) that are willing to share data with each other to pool their individual-level patient data into a single repository. The data repository is usually hosted by one medical site or by an external third party (e.g., a cloud provider), playing the trusted dealer role. The main advantage of this model is that the trusted dealer enables authorized investigators to access all the patient-level information needed for data cleaning and for conducting statistical analysis. Moreover, such a data-sharing model minimizes infrastructure costs at medical sites as data storage and computation are outsourced. However, from a data privacy perspective the centralized model is often difficult to realize, especially when medical and genetic data have to be exchanged across different jurisdictions. The central site hosting the data repository represents a single point of failure in the data-sharing process. All participating sites have to trust such a single entity for protecting their patient-level data [7].

To minimize sensitive information leakage from data breaches, traditional anonymization techniques include suppressing of directly identifying attributes, as well as the generalizing, aggregating or randomizing quasi-identifying attributes in individual patient records. In particular, the k-*anonymity* privacy model,[8] is a well-established privacy-preserving model that aims to reduce the likelihood of re-identification attacks "singling out" an individual. Specifically, the k-anonymity model ensures that for each combination of quasi (or indirect) identifiers there exist at least *k* individuals who share the same attributes.

However, given increased sophistication of re-identification attacks [8–14] and the rising dimensionality (number of clinical and genetic attributes) of patient data, the above-mentioned countermeasures are inadequate to ensure a proper level of anonymization and preserve acceptable data utility. As a result, these conventional anonymization techniques for individual-level patient data are rarely used in practice. Researchers prefer to rely on simple pseudonymization techniques (such as replacing direct identifiers with pseudonymous codes) combined with legal measures defining each party's responsibilities regarding data transfer, access and use. This process generates administrative overheads that slow down the pace of biomedical research. Further, although designed to comply with data protection regulations, contractual safeguards may not eliminate the risk of individuals being reidentified [15]. As we argue below, combining traditional pseudonymization mechanisms and governance strategies meet the legal standard of pseudonymization but not anonymization under the GDPR.

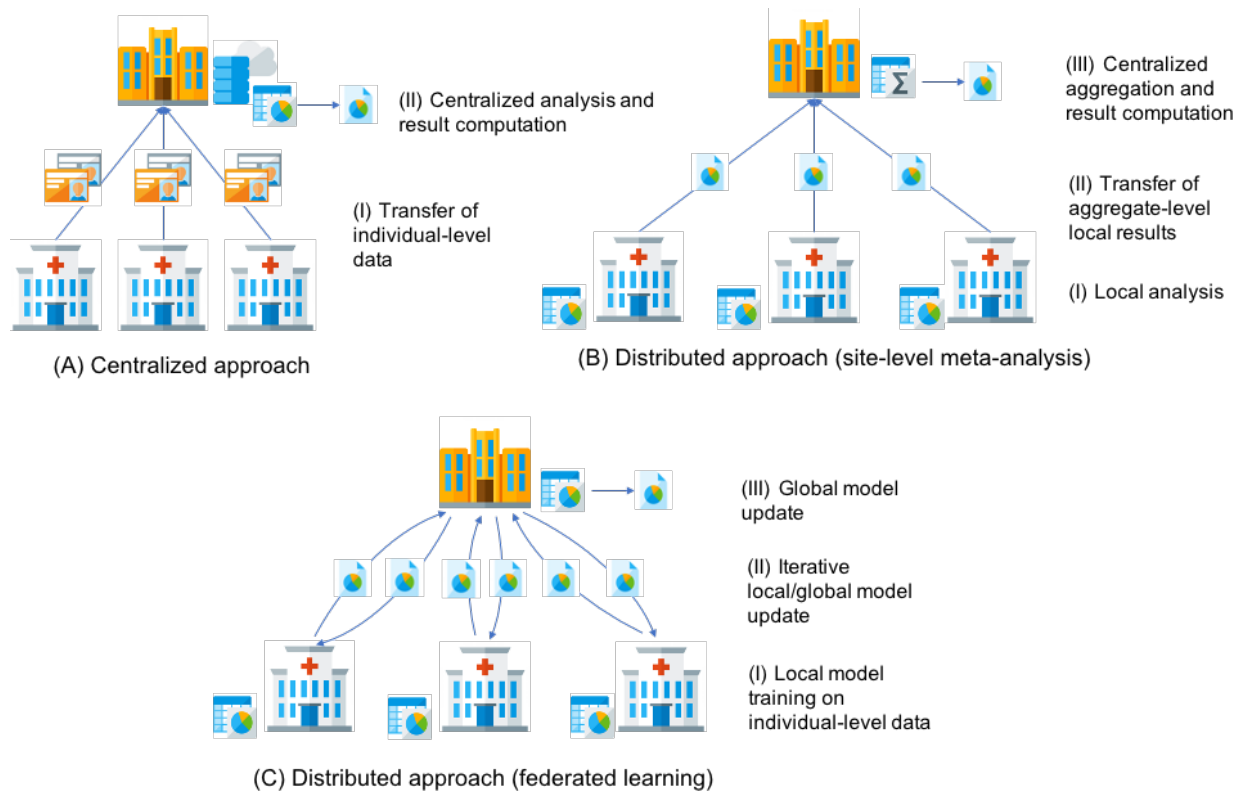

*Figure 1. Overview of the three main data-sharing models: (A) Centralized, (B) Decentralized (site-level meta-analysis), (C) Decentralized (federated learning).*

**Decentralized model: site-level meta-analysis**
As opposed to the centralized data-sharing model, the decentralized model does not require patient-level data to be physically transferred out of the medical sites' IT infrastructure. Medical sites keep control over their individual patient-level patient data and define their own data governance principles. For each clinical study, the statistical analysis is first computed on local datasets. The resulting local statistics are then sent to the site responsible for the final meta-analysis that aggregates the separate contribution of each data provider [16] to obtain the final result of the analysis. Under this model, the site performing the meta-analysis is trusted by all other sites for the protection of their local statistics. As local statistics have a significantly lower dimensionality with respect to individual-level data, there is a lower risk of re-identification in the decentralized data-sharing model.

However, the sharing of only aggregate level data does not guarantee patients' privacy by itself. Some aggregate-level statistics may be too low for certain subpopulations (such as patients with rare diseases) and can be considered personally identifying. Moreover, in some circumstances aggregate-level data from local analyses can be exploited to detect the presence of target individuals in the original dataset. For example, an attacker may already hold the individual level data of one or several target individuals [17–21]. This membership information can be subsequently used to infer sensitive and sometimes stigmatizing attributes of the target individuals. For example, detecting the membership of an individual in a HIV-positive cohort reveals her HIV status. The intuition behind these attacks is to measure the similarity between the individual-level target data with statistics computed from the study dataset and statistics computed from the general population. The attacker's certainty about the target's membership in the dataset increases with the similarity of the target's data to the statistics derived from the study dataset.

To address these inference attacks, clinical sites can anonymize their local statistics by applying obfuscation techniques that mainly consist in adding a certain amount of statistical noise on the aggregate-level data before transfer to third parties. This process enables data providers to achieve formal notions of privacy such as *differential privacy* [22,23]. In the statistical privacy community, differential privacy is currently considered as guaranteeing the likelihood of re-identification from the release of aggregate-level statistics can be minimized to a quantifiable value. Similar to anonymization techniques for individual-level data, statistical obfuscation techniques degrade the utility of aggregate-level data. Consequently, the amount of noise introduced by data obfuscation has to be carefully calibrated to reach the desired compromise between utility and privacy. Often, when each data provider adds the required amount of noise to reach an acceptable level of privacy, the resulting aggregated results stemming from a meta-analysis are too distorted to be reliable [24].

Beyond privacy considerations, this approach also suffers from a lack of flexibility as the medical sites involved in the analysis have to coordinate before the analysis on the choice of parameters and covariates to be considered. This coordination often depends on manual approval, impeding the pace of the analysis itself. Finally, as opposed to the centralized approach, accuracy of results from a meta-analysis that combines the summary statistics or results of local analysis can be affected by cross-study heterogeneity. This can lead to inaccurate and misleading conclusions [25].

**<u>Decentralized model: federated analysis and learning</u>**
The federated model is an evolution of the decentralized model based on site-level meta-analysis. Instead of sharing the results of local analyses, the participating data providers collaborate to perform a joint analysis or the training of a machine learning model in an interactive and iterative manner, only sharing updates of the model's parameters. One of the medical sites participating in the multi-centric research project (typically the site responsible for the statistical analysis) becomes the reference site (or central site) and defines the model/analysis to be trained and executed on the data distributed across the network. This model is referred to as the "global" model. Each participating site is given a copy of the model to train on their own individual-level data. Once the model has been trained locally for a number of iterations, the sites send only their updated version of the model parameters (aggregate-level information) to the central site and keep their individual-level data at their premises. The central site aggregates the contributions from all the sites and updates the global model [26]. Finally, the updated parameters of the global model are shared again with the other sites. The process repeats iteratively till convergence of the global model.

With respect to the distributed data-sharing approach based on site-level meta-analysis, this federated approach is more robust against heterogeneous distributions of the data across different sites, thus yielding results accuracy that is comparable to the results obtained with the same analysis conducted using the centralized model. Moreover, this approach does not suffer from the loss in statistical power of conventional meta-analyses. Prominent projects that have attempted to employ federated approaches to analysis and sharing of biomedical data are the DataSHIELD project [27] and the Medical Informatics Platform of the Human Brain Project [28].

The federated data-sharing approach combines the best features of the other two approaches. However, although the risk or re-identification is reduced compared to the centralized approach, the federated approach remains vulnerable to the same inference attacks of the meta-analysis approach. These inference attacks exploit aggregate-level data released during collaboration [29–32]. The potential for an inference attack is even increased compared to a meta-analysis-based approach. This is due to the iterative and collaborative nature of the data processing, allowing adversaries to observe model changes over time and with specific model updates. Melis et al [33] show that updates of model parameters transferred during the collaborative training phase can be used to infer the membership of a target individual in the training datasets as well as some properties associated with a particular subset of the training data. This inference is possible if the context of the data release enables the attacker to easily access some auxiliary individual-level information about the target individual. In legal terms (as discussed below), these aggregate-level data can potentially be considered personal data. As for the meta-analysis approach, the use of obfuscation techniques can be used to anonymize model's updates at each iteration. Nevertheless, the required perturbation can severely affect the performance of the final model [24].

Finally, regardless of the type of distributed data-sharing model, obfuscation techniques for anonymizing aggregate-level data are rarely used in practice in medical research because of their impact on data utility. As a result, these technical privacy limitations are usually addressed via additional legal and organizational mechanisms. For the DataSHIELD project, access is limited to organizations that have consented to the terms of use for DataSHIELD and have sought appropriate ethics approval to participate in a DataSHIELD analysis [34]. Therefore, implementing the platform will require cooperating with governments and institutions so they are comfortable with exposing sensitive data to the platform [27]. However, as we discuss below, advanced PETs can also guarantee data privacy.

## Minimizing risks by leveraging advanced PETs

In the last few years, a number of cryptographic PETs have emerged as significant potential advances for addressing the above-mentioned data protection challenges that still affect medical data sharing in the decentralized model. Although hardware-based approaches could be envisioned for this purpose, they are usually tailored to centralized scenarios and introduce a different trust model involving the hardware provider. Further, they also depend on the validity of the assumptions on the security of the hardware platform, for which new vulnerabilities are constantly being discovered. In this paper, we focus on two of the most powerful software-based

PETs *Homomorphic Encryption* and *Secure Multiparty Computation*. Both rely on mathematically proven guarantees for data confidentiality, respectively grounded on cryptographic hard problems and non-collusion assumptions.

**Homomorphic Encryption**
Homomorphic Encryption (HE) [35] is a special type of encryption that supports computation on encrypted data (ciphertexts) without decryption. Thanks to this property, homomorphically encrypted data can be securely handed out to third parties, who can perform meaningful operations on them without learning anything about their content. Fully homomorphic encryption schemes, or schemes enabling arbitrary computations on ciphertexts, are still considered non-viable due to the high computational and storage overheads they introduce. Current practical schemes that enable only a limited number of computations on ciphertexts (such as polynomial operations) have reached a level of maturity that permits their use in real scenarios.

**Secure Multiparty Computation**
Secure Multiparty Computation (SMC) [36–40] protocols enable multiple parties to jointly compute functions over their private inputs without disclosing to the other parties more information about their inputs than what can be inferred from the output of the computation. This class of protocols is particularly attractive in privacy-preserving distributed analytic platforms due to the great variety of secure computations they enable. However, this flexibility includes a number of drawbacks that hinder their adoption, including high network overhead and parties required to be online during computation.

**Multiparty Homomorphic Encryption**
The combination of SMC and HE was proposed to overcome their respective overheads and technical limitations; we refer to it as Multiparty Homomorphic Encryption (MHE) [41–44]. MHE enables flexible secure processing by efficiently transitioning between encrypted local computation, performed with HE, and interactive protocols (SMC). It can be used to choose the most efficient approach for each step within a given workflow, leveraging the properties of one technique to avoid the bottlenecks of the other. Moreover, MHE ensures that the secret key of the underlying HE scheme never exists in full. Instead, it distributes the control over the decryption process across all participating sites, each one holding a fragment of the key. All participating sites have to agree to enable the decryption of any piece of data, and no single entity alone can decrypt the data.

Unlike HE or SMC alone, MHE provides effective, scalable and practical solutions for addressing the privacy-preserving issues that affect the distributed/federated approach for data sharing. For example, systems such as Helen,[45] MedCo,[46] or POSEIDON [47] use MHE to guarantee that all the information interchanged between the sites is always in encrypted form, including aggregate data such as model parameters and model updates, and only the final result (the computed model or the predictions based on this model) is revealed to the authorized user. Finally, MHE reduces the need of obfuscation techniques to protect aggregate-level data from inference attacks. Further, data utility, which is typically lost with privacy-preserving distributed approaches that only rely on obfuscation techniques, can be significantly improved. As aggregate-level data transfer and processing across participating sites during the analysis or training phase remains always encrypted, obfuscation can be applied only to the decrypted final result of the analysis that is released to the data analyst, instead of being applied to all local model updates at each iteration. Hence, MHE enables a much lower utility degradation for the same level of re-identification risk.

# Regulatory hurdles for the use of encryption technologies

In this section we focus on the features of EU data protection law concerning encryption and data sharing. We focus on the GDPR because of the persistence of national divergences in member state law, despite the passage of the GDPR. In particular, the GDPR provides member states can introduce further restrictions on processing of genetic data, biometric data or health related data. This flexibility increases the potential for divergences in national law that require customized contracts between institutions in different EU member states [4].

**Data anonymization and pseudonymization**
The GDPR defines personal data as concerning an "identifiable natural person". Therefore, pseudonymized data, where all identifiers have been removed from that data, remains personal data. However, the provisions of the GDPR do not concern anonymized data, or data which has been processed so individuals are no longer identifiable. In particular, anonymized data may be used for "research or statistical processing" without the need to comply with the GDPR.

Spindler and Schmechel [48] note there are two conflicting approaches to classifying personal and anonymized data. The first is an absolute approach, where anonymized data constitute personal data if there is even a theoretical chance of re-identification. This approach represents the state of national law in a minority of EU member states, such as France [49]. The second is the relative approach, where anonymized data is no longer personal data if it is reasonably likely that methods do not exist to re-identify individuals [48]. This approach represents the state of national law in countries such as Ireland, where the Irish Data Protection Commission has held that data are anonymized if it is unlikely current technology can reidentify that data [50]. Likewise, the German Federal Ministry for Economic Affairs and Energy held that data (including health related personal data) is anonymized under the BDSG where individuals cannot be reidentified with reasonable effort [51]. In both of these jurisdictions, if an unreasonable effort were required to re-identify anonymized data, then it would no longer be personal data [48].

At the supranational level, the former Article 29 Working Party (now the European Data Protection Board) has favored a relative over an absolute approach to anonymization. First, the Article 29 Working Party held that the words "means reasonably likely" suggests a theoretical possibility of re-identification will not be enough to render that data personal data [52]. A subsequent opinion of the Working Party reinforced this support for the relative approach and compared different techniques for anonymization or pseudonymization. For example, encrypting data with a secret key means that data could be decrypted by the key holder. For this party, the data would therefore be pseudonymized data. But if a party does not have the key, the data would be anonymized.

Likewise, if data is aggregated to a sufficiently high level, this data would no longer be personal data [53]. Nevertheless, following the Article 29 Working Party's ruling, no single anonymization technique can fully guard against orthogonal risks of re-identification [54].

**Data Processing:**
The GDPR's provisions apply to data controllers, or entities determining the purpose and means of processing personal data. This definition encompasses both healthcare institutions and research institutions. Data controllers must guarantee personal data processing is lawful, proportionate, and protects the rights of data subjects. In particular, the GDPR provides that encryption should be used as a safeguard when personal data is processed for a purpose other than which it was collected. Although the GDPR does not define encryption, the Article 29 Working Party treats encryption as equivalent to stripping identifiers from personal data. The GDPR also lists encryption as a strategy that can guarantee personal data security. Further, the GDPR emphasizes that data controllers should consider the "state of the art", along with the risks associated with processing, when adopting security measures. The GDPR also provides that data processing for scientific purposes should follow the principle of data minimization. This principle requires data processors and controllers to use non-personal data unless the research can only be completed with personal data. If personal data is required to complete the research, pseudonymized or aggregate data should be used instead of directly identifying data.

The GDPR imposes obligations on data controllers with respect to the transfer of data, particularly outside of the EU. Specifically, the GDPR requires the recipient jurisdiction to offer adequate privacy protection before a data controller transfers data there. Otherwise, the data controller must ensure there are organizational safeguards in place to ensure the data receives GDPR equivalent protection. Further, data controllers must consider the consequences of exchanging data between institutions, and whether these are joint controllership or controller-processor arrangements. Under the GDPR, data subject rights can be exercised against any and each controller in a joint controllership agreement. Further, controllers must have in place an agreement setting out the terms of processing. By contrast, a data controller-processor relationship exists where a controller directs a data processor to perform processing on behalf of the controller, such as a cloud services provider. The GDPR provides that any processing contract must define the subject matter, duration, and purpose of processing. Contracts should also define the types of personal data processed and require processors to guarantee both the confidentiality and security of processing.

## Advanced PETs and EU data governance requirements

In this section, we argue that MHE, or HE and SMC used in concert, meets the requirements for anonymization of data under the GDPR. Further, we argue the use of MHE can significantly reduce the need for custom contracts to govern data sharing between institutions. We focus on genetic and clinical data sharing due to the potential for national derogations pertaining to the processing of health-related data. Nevertheless, our conclusions regarding the technical and legal requirements for data sharing using MHE, or HE and SMC may apply to other sectors, depending on regulatory requirements [55].

Under the GDPR, separating pseudonymized data and identifiers is analogous to separating decryption keys and encrypted data. To this end, Spindler and Schmechel suggest that encrypted data remain personal data to the entity holding the decryption keys [48]. The encrypted data also remain personal data for any third party with lawful means to access the decryption keys. Applying this approach to HE, if a party has access to the decryption key corresponding to the encryption key that was used to homomorphically encrypt data, that party will have access to personal data. Likewise, if a party has lawful access to data jointly processed as part of SMC, that data will remain personal data for that party [56].

Whether a party to data processing using advanced PETs has lawful access to data or decryption keys depends on the legal relationship between the parties. With respect to joint controllership, recent CJEU case law has established that parties can be joint controllers even without access to personal data (Case C-210/16; Case C-25/17; Case C-40/17). In Case C-210/16, the CJEU held that the administrator of a fan page hosted on Facebook was a joint controller despite only having access to aggregate data (Case C-210/16, [para. 38]). However, Article 26 para. 1 requires joint controllers to establish a contract allocating responsibility for processing of personal data. Hospitals or research institutions processing patient data using SMC jointly determine how these data are processed. These entities would be classified as joint controllers, at least when engaging in secret sharing (as a joint purpose of data processing). These entities would need an agreement to establish that only the entity with physical access to patient data can access that data. If a request is made to a hospital or research institution that does not possess this data, the request must be referred to the entity that does.

Applying these principles to processing with PETs, for HE, there is no mathematical possibility of decrypting the data without the decryption key. This holds true when both the data are at rest or processed in the encrypted space via secure operations such as homomorphic addition or multiplication. Whether data processed as part of SMC or MHE remains personal data depends on whether entities have lawful access to personal data or decryption keys respectively. If entities can only access personal data they physically hold as part of a joint controller agreement, the data fragments exchanged during secret sharing via SMC are not personal data. Likewise, under MHE each individual entity only has access to a fragment of the decryption key, which can only be recombined with the approval of all other entities holding the remaining fragments. This argument is reinforced by Recital 57 of the GDPR, which provides controllers forbidden from identifying individuals are not required to collect identifying information to comply with the GDPR.

Therefore, we submit that both HE and SMC, when used alone or together through MHE can jointly compute health related data whilst complying with the GDPR. These data remain anonymous even though entities processing data using MHE are joint controllers. Further, the use of advanced PETs should become a best standard for the processing of health-related data for three reasons. Firstly, the Article 29 Working Party has recommended using encryption and anonymization techniques in concert to protect against orthogonal privacy risks and overcome the limits of individual techniques. Secondly, the GDPR emphasizes the use of "state of the art" techniques for guaranteeing the processing of sensitive data. HE, SMC, and MHE are considered state of the art technologies in that they carry a mathematical guarantee of privacy. Thirdly, the Article 29 Working Party has held the data controller is responsible for demonstrating that the data has been and remains anonymized. Further support from this argument comes from a case heard before the Swiss Federal Supreme Court, A_365/2017. In this case, the Federal Supreme Court endorsed a relative approach to anonymization, but also placed the onus on the data controller to establish anonymization (A_365/2017, para. 5.12). Switzerland is not a member of the EU and does not have to comply with the GDPR. However, Switzerland's close proximity to the EU means the Federal Act on Data Protection (FADP) has been revised. These revisions ensure the continued free exchange of data between Switzerland and EU countries [57].

Therefore, we argue that MHE involves processing anonymized data under EU data protection law. Although HE, SMC and MHE do not obliviate with the need for a joint controllership agreement, they lessen the administrative burden required for data sharing. Further, they promote the use of standard processing agreements that can help ameliorate the impacts of national differences within and outside the EU. Accordingly, we submit that MHE, along with other forms of advanced PETs, should represent the standard for health-data processing in low trust environments [58]. This processing can include performing computations on sensitive forms of data, such as providing genomic diagnoses without revealing the entire sequence for a patient [59]. Further, the encrypted outputs of HE and secure multiparty computation are mathematically private, as they do not reveal any personal data [60]. The various states of data processed using novel PETs such as MHE is displayed in Figure 2 below.

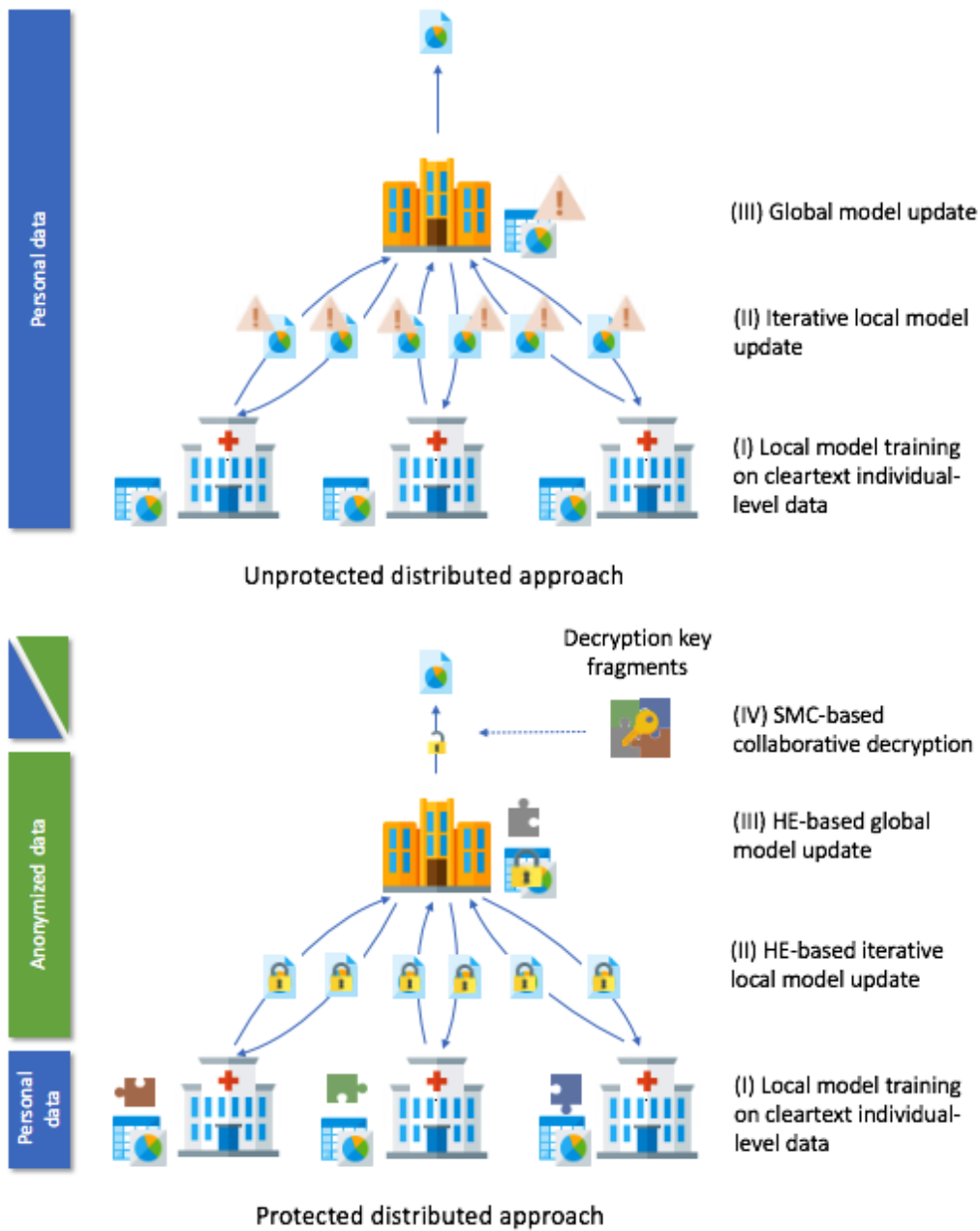

*Figure 2.* Comparison of the status of personal data under a distributed approach relying on traditional PETs (e.g., aggregation and pseudonymization) and a distributed approach relying on MHE (e.g., HE and SMC).

Further, Table 1 demonstrates the status of personal data at different stages of processing.

| Scenario | Status of data based on the scenario |
|---|---|
| **Scenario A:** Hospital/research institution | Personal data |
| physically holds personal data | |
| **Scenario B:** Hospital/research institution has legal access to decryption key/personal data | Pseudonymized data |
| **Scenario C:** Hospital/research institution combine decryption keys/personal data to process data | Anonymized data |
| **Scenario D:** Third party (cloud service provider) carries out processing, hospitals share encryption keys jointly | Anonymized data |

Table 1: The status of data when held by entities at different stages of processing

The lack of reliance on custom contracts may encourage institutions to align their data formats to common international interoperability standards. In the next section, we turn to address the standardization of these advanced PETs.

## Regulatory instruments to encourage the use of novel PETs

At present, regulatory instruments provide limited guidance on the different types of PETs required to process medical data in a privacy conscious fashion. However, the techniques described in this paper may represent a future best standard for processing medical data for clinical or research purposes. Because of the recency of both technologies, the standardization of homomorphic encryption and secure multiparty computation is ongoing, with the first community standard released in 2018 by HomomorphicEncryption.org [61].

Further, there are numerous documents published by data protection agencies that can aid the development of such guidelines. For example, the French Data Protection Agency, *Commission Nationale Informatique & Libertes* (CNIL), published a set of guidelines following the passage of the GDPR on how to secure personal data. This document provides recommendations on when encryption should be used, including for data transfer and storage [62]. Likewise, the Spanish Data Protection Agency has already recommended using homomorphic encryption as a mechanism for achieving data privacy by design pursuant to Article 25 of the GDPR [63].

Nevertheless, any standards will need to be continually updated to respond to new technological changes. For example, one of the most significant drawbacks of FHE is the complexity of computation. This computational complexity makes it hard to predict running times, particularly for low powered devices such as wearables and smartphones. For the foreseeable future, this may limit the devices upon which fully homomorphic encryption can be used [64]. Therefore, specialized standards may need to be developed for using homomorphic encryption on low powered devices in a medical context. Although homomorphic encryption and secure multiparty computation offer privacy guarantees, there is still an orthogonal risk of reidentifying individuals from aggregate-level results that are eventually decrypted and can be exploited by inference attacks [17,19,20,65]. However, as mentioned earlier, the use of MHE or SMC enables the application of statistical obfuscation techniques for anonymizing aggregate-level results with a better privacy-utility trade-off than the traditional distributed approach, thus facilitating the implementation of end-to-end anonymized data workflows.

A final consideration relates to ethical issues that exist beyond whether HE, MPC, and MHE involve processing anonymized or personal data. First, the act of encrypting personal data constitutes further processing of that data under data protection law. Therefore, healthcare and research institutions must seek informed consent from patients or research subjects [48]. Institutions must consider how to explain these technologies in a manner that is understandable and enables the patient to exercise their rights under data protection law. Secondly, the institution that holds the data must have procedures in place that govern who can access data encrypted using advanced PETs. Institutions should also determine which internal entity is responsible for governing access requests. These entities can include ethics review committees or data access committees [2].

## Conclusion

Medical data sharing is essential for modern clinical practice and medical research. However, traditional privacy-preserving technologies based on data perturbation, along with centralized and decentralized data sharing models, carry inherent privacy risks and may have high impact on data utility. These shortcomings mean that research and healthcare institutions combine these traditional privacy-preserving technologies with contractual mechanisms to govern data sharing and comply with data protection laws. These contractual mechanisms are context-dependent and require trusted environments between research and healthcare institutions. Although federated learning models can help alleviate these risks as only aggregate-level data are shared across institutions, there are still orthogonal risks to privacy from indirect re-identification of patients from partial results [58]. Further, changes in case law (such as the already mentioned recent invalidation of the US-EU Privacy Shield under *Schrems II*) can undermine data sharing with research partners outside the EU. In this paper, we demonstrated how these privacy risks can be addressed through the use of Multiparty Homomorphic Encryption (MHE), an efficient combination of homomorphic encryption and secure multiparty computation. In particular, we demonstrated how homomorphic encryption and secure multiparty computation can be used to compute accurate federated analytics without needing to transfer personal data. Combining these technologies (Multiparty Homomorphic Encryption) for medical data sharing can improve the performance overheads of privacy enhancing technology whilst reducing the risk of GDPR non-compliance. Further, personal data does not leave the host institution where it is stored when processed using MHE. Therefore, the lack of personal data transfer with MHE will encourage increased data sharing and standardization between institutions. Data protection agencies, as well as healthcare and research institutions, should promote MHE and other advanced PETs for their use to become widespread for clinical and research data sharing.

# Acknowledgments

We are indebted to Dan Bogdanov, Brad Malin and Sylvain Métille for their invaluable feedback on earlier versions of this manuscript. This work was partially funded by grant #2017-201 (Project "Data Protection and Personalized Health") of the Personalized Health and Related Technologies Program supported by the Council of the Swiss Federal Institutes of Technology.